# Diagnosis of electron density and temperature by using collisional radiative model in capacitively coupled Ar plasmas I : triple-frequency discharges

Hao Zheng[1], Jidun Wu[1], Qilu Cao[1], Jiaojiao Zhang[1] and Xiaojiang Huang[1,2,3]

[1] College of Science, Donghua University, Shanghai 201620, China
[2] Member of Magnetic Confinement Fusion Research Centre, Ministry of Education of China, Shanghai 201620, China
[3] Textiles Key Laboratory for Advanced Plasma Technology and Application, Shanghai 201620, China

E-mail: huangxj@dhu.edu.cn

**Abstract**

An 18-level argon collisional radiative model (CRM) suitable for low pressure was established. The model can be solved by combining the optical emission spectroscopy (OES) with Langmuir probe calibration. In the capacitively coupled plasmas (CCPs) with different frequency and power, the electron temperature and density obtained by the model were compared with those measured by Langmuir probe. It is found that the calibration point at any frequency or power is suitable for the fixed pressure. This method was then applied to the diagnosis of triple-frequency (TF) CCPs, it is shown that the high frequency (HF) power has the greatest influence on electron density and electron temperature, that of low frequency (LF) power is the least, and the intermediate frequency (IF) power was between the two. Compared with the dual-frequency (DF) CCPs, it is found that with the increase of IF power, the HF power can control the electron density more independently with less influence on the electron temperature.

## 1. Introduction

In the past few decades, capacitively coupled plasmas (CCPs) have been extensively applied to the semiconductor industry, such as plasma etching and thin film deposition [1]. In order to control the flux and energy of ions independently in CCPs, many numerical simulations and experiments have been made [2-5]. Those studies found that it could be achieved by using dual-frequency (DF) discharges, which were driven by one high and one low RF power sources at the same time. In DF-CCPs, HF mainly controls plasma density, while LF mainly controls ion energy. However, little attention has been paid to the effect of high and low frequency on electron temperature.

Since the addition of a radio frequency power source can make CCPs more regulate, some studies on TF-CCPs have been done in recent years [6-9]. Wu et al [6] developed an analytical model to predict the ion energy distribution functions (IEDFs) in dual and triple frequency plasma discharges. The results of this model were well consistent with PIC simulations. They found that the peaks of middle energy became evident with the addition of IF drive to a DF drive. Kawamura *et al* [7] investigated the influence of HF on the plasma uniformity in TF-CCPs using a fast 2D fluid-analytical simulation. They observed that the addition of a HF source appeared to effectively improve the radial plasma uniformity. With a combination of an analytical model and PIC simulations, Sharma *et al* [8] demonstrated the IEDFs at the electrode could be controlled by tuning the IF. Chen *et al* [9] presented a fast semi-analytical method to study the role of HF current density in TF-CCPs. The averaged ion energy and energy spread of the IEDFs decreased with the increase of HF current density. Nevertheless, these works were mainly based on analytical models or simulations, few researchers have studied the function of IF in TF discharges experimentally.

The experimental study of CCPs replies on various diagnostic techniques. Langmuir probe and spectrometer are the common plasma diagnostic methods. The single probe can be used to measure the electron temperature and density conveniently. However, the single probe is interfered by the RF power and needs to be filtered. The interference will increase sharply, and filtering is very difficult in Multi-frequency discharges. Therefore, the single probe method cannot be applied to discharges with multiple RF sources [10]. Johnson and Malter [11] developed a symmetrical dual probe diagnostic system, which was successfully used to diagnose complex plasma discharges, because it does not require a special electrode as a reference potential. Although the dual probe can effectively shield the interference of RF power, the dual probe can only respond to the high-energy electrons in the plasma, and is not suitable for the low-voltage and low-power discharge environment [10]. Piejak *et al* [12] and Stenzel [13] presented and developed a floating hairpin probe diagnostic technique. The electron density has been simply determined through the shift of microwave resonant frequency in plasma with respect to that in vacuum. However, a value of electron temperature must be assumed to further obtain the electron density using this approach.

Optical emission spectroscopy (OES) is another common plasma diagnostic method. The OES is a non-intrusive diagnostic technique and not affected by RF electric field, magnetic field and high electric potential of plasma [14]. Therefore, OES diagnosis is a common method for diagnosing multi-frequency discharges. However, some modeling methods are needed to obtain useful information from spectral line intensities. The electron excitation temperature can be calculated by using the line intensity ratio, but there is a certain difference between the electron excitation temperature and the actual electron temperature in plasma. The Stark stretch, which calculates the electron density, works only for high density plasmas. Those two methods can only obtain electron temperature or electron density separately. The CRM is based on a model containing the main kinetic processes of excited particles to calculate the relationship between emission spectra and plasma parameters. For each type of particle, a rate-equilibrium equation is generally formulated, which includes not only the space-time evolution of the particle, but also the various collision reactions and radiation processes involved. Theoretically, the electron density and temperature can be calculated simultaneously, but the solution is complicated [15-17].

In the past several decades, CRMs have been widely developed for analyzing Ar plasma discharges. Vlček *et al* [18] built a CRM applicable to argon plasma discharges with an extensive range of conditions, and 65 effective levels were considered in this model. Bultel *et al* [19] developed a nonlinear time-dependent CRM for recombining argon and discussed the influence of Ar2+. They found that Ar2+ might cause the increment of the time to reach a quasi-steady-state by a factor of 100. Palmero *et al* [17] presented a simple CRM with 12 effective levels for low-pressure Ar discharges and found that the lower lying excited states of argon played an important role in the excitation. The model was used to calculate the electron density and electron temperature in an argon magnetron sputtering plasma produced at different electromagnetic powers and gas pressures, as a function of the intensity of the optical emission lines. Zhu *et al* [20] established a CRM of Ar applicable from low pressure to normal pressure, mainly studying the change of kinetic energy of 1s and 2p level particles, which also considered the influence of particles and ions on argon discharge. Recently, Kovalev *et al* [21] also proposed a zero-dimensional CRM to calculate the 1s states densities of Ar. They observed that the calculated densities of resonance states were underestimated, while the calculated metastable densities were well consistent with the experimentally measured one. Most of the researches on the CRM mainly focused on the model itself, and our main purpose in this work is to use the model for experimental diagnosis.

In this work, an Ar 18-level CRM was developed for low-pressure argon discharges. Then the results of electron temperature and density were compared using OES combined with CRM method with the results of the probe method. After that, OES combined with CRM method was used to study the electron temperature and density with high, intermediate and low Rf power in TF-CCPs. At the same time, comparing with the DF discharge, the influence of the IF power on the plasma in the TF discharges was analyzed.

The paper is structured as follows. In Section 2, the experimental setup and diagnostic systems are introduced briefly. In Section 3, the CRM and solving method is described. In Section 4, results are shown and discussed. Conclusions are drawn in Section 5.

## 2. Experimental setup

Figure 1 shows the schematic diagram of the CCP reactor and the diagnostic system. The diameter of stainless-steel reactor is 200 mm and the height of it is 180 mm. There two aluminum round plate electrodes are 50 mm in diameter and placed parallel in the reactor. The gap between two electrodes is fixed to 30 mm. In the single-frequency (SF) discharges, a RF power source is connected to the bottom electrode via a matching box, and the top electrode is grounded. In the TF discharges, two RF power sources at 13.56 MHz and 2 MHz are connected to the top electrode via respective matching box and a homemade frequency mixer. The 27.12 MHz power is applied to the bottom electrode through a matching box. Note that the power mentioned in this work is just the applied RF power. Two electrodes are cooled by water during discharges. With a 500 r/s turbomolecular pump and a mechanical pump, the base pressure of the chamber is as low as $10^{-3}$ Pa. Pure argon (99.995%) is controlled at 10 sccm into the chamber by a mass flow controller, and the pressure in the chamber is fixed at a low pressure of 1.5 Pa.

An OES (Avaspec-2048TEC) in the absolute irradiance mode is used to monitor the optical emission through a lens and optical fiber along the midplane from the CCP discharges, with its spectral resolution of 0.13 nm, and the range of wavelength from 200 nm to 900 nm. The Langmuir probe system (MMLAB-probe-1) consists of probe, filter, driver and data acquisition. The two diagnostic systems are connected to respective side window of the discharge chamber, while keeping the probe at the same height as the lens, in the middle of the two electrodes.

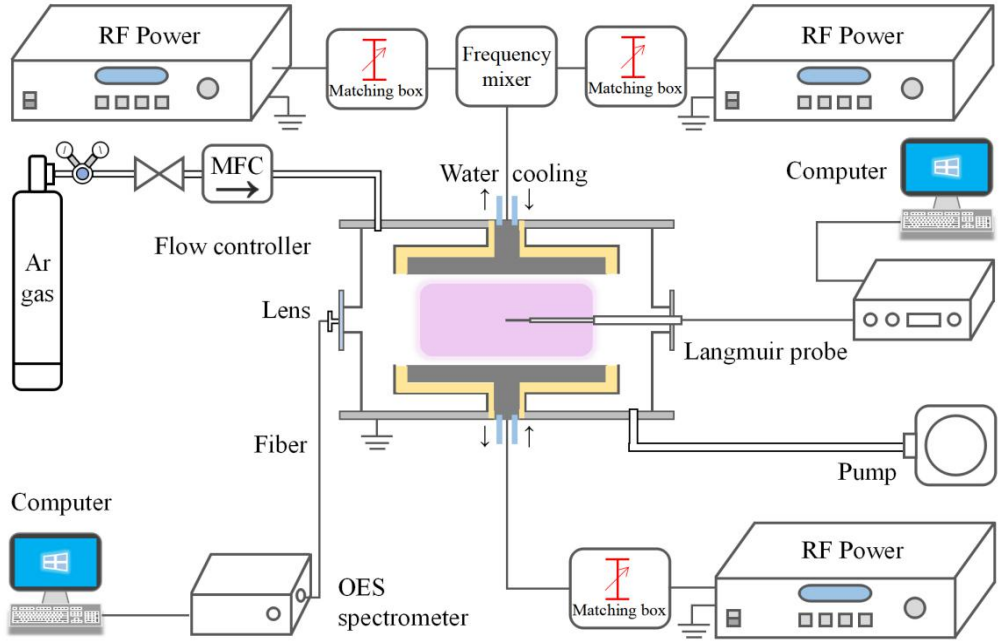

**Figure 1.** Schematic of the CCP reactor and diagnostic system.

## 3. Argon collision radiation model

Our model is a simplified version of the model in [20]. In the low-pressure (< 100 Pa) nonequilibrium regime, the dominant processes are the ground-state and metastable-level electron-impact excitation, and radiation to the ground state or 1s levels. Therefore, the lower excited states of argon are mainly considered. The model is an Ar 18-level CRM, in which individual the ground level, all 4 levels of 1s and 10 levels of 2p are used instead of the effective levels. Also, it includes three effective levels for higher excited states of argon, which are 2s3d, 3p and high level (hl) respectively.

18-level is set as level 1 = ground-state, level 2-5 = $1s_5$-$1s_2$, level 6-15 = $2p_{10}$-$2p_1$, level 16 = 2s3d, level 17 = 3p, and level 18 = hl. We use the normalized populations $x_i = n_i/n_0$, where, $n_i$ is a population density of level i, $n_0$ is the total number of argon atoms in the plasma. Then the rate balance equation for each kind of excited argon atoms (level i from 2 to 18) is

$$n_e \sum_{j=1,j\neq i}^{18} Q_{j\to i} x_j + \sum_{j>i} \Gamma_{j\to i} A_{j\to i} x_j - n_e \sum_{j=1,j\neq i}^{18} Q_{i\to j} x_i - n_e Q_{i\to ioniz} x_i - \sum_{j<i} \Gamma_{i\to j} A_{i\to j} x_i - K_i x_i = 0 \quad (1)$$

and

$$\sum_{i=1}^{18} x_i = 1 \quad (2)$$

Where, $Q$ as the electron-impact reaction rate coefficient is a function of electron temperature ($T_e$), $n_e$ is electron density, $A$ is the Einstein coefficient, $\varGamma$ is the volume-averaged escape factor, and $K$ refers to the wall reaction coefficient. The parameters of $Q$ and $A$ referenced in [20] of Zhu et al. are adopted. The parameters of $\varGamma$ and $K$ are adopted in [22] and [23], respectively. In equation (1), term (1) refers to the population of electron-impact excitation from other levels to level i. Term (2) is radiation from upper levels to level i. Term (3) is electron-impact losses from level i to other levels. Term (4) is also electron-impact losses for the ionization of level i. Term (5) is the losses due to level i radiation to lower levels. Term (6) refers to the losses due to the wall reaction.

Equations (1) and (2) including 18 equations with 20 unknown quantities, which are $x_1, x_2...x_{18}$ and $T_e$, $n_e$. It means that if two unknown quantities can be measured, the rests can be calculated. The emission lines with wavelength 696.5nm ($2p_2$-$1s_5$) and 750.4 nm ($2p_1$-$1s_2$) are mainly emitting from the level 14 and level 15, respectively. The relations of the line strength of 750.4 nm ($I_{750}$) and 696.5 nm ($I_{696}$) to the corresponding excited state population are solved using the method of [17]. When the gas pressure is fixed, an approximation relation can be written

$$I_{696} = C_{696} \cdot x_{14} , I_{750} = C_{750} \cdot x_{15} \qquad (3)$$

where, $C_{750}$ and $C_{696}$ are the calibration constants. They can be obtained by an absolute calibration with the Langmuir probe. Firstly, one discharge condition is selected as the calibration point, and $T_e$, $n_e$ are measured by the Langmuir probe and $I_{750}$, $I_{696}$ measured by OES simultaneously. Then $T_e$ and $n_e$ are taken as input parameters in the model, equations (1) and (2) become linear equations, and $x_{14}$, $x_{15}$ can be obtained through solving them. With $x_{14}$, $x_{15}$ values and $I_{750}$, $I_{696}$, $C_{750}$ and $C_{696}$ are calculated by using equations (3). Therefore, as knowing the calibration constants $C_{750}$ and $C_{696}$, $I_{750}$ and $I_{696}$ in other discharge conditions can be used as input quantities in the model. Here, equations (1) and (2) become nonlinear equations, and we solve the equations by CR, a homemade visual program we compiled with Python 3, to obtain $T_e$ and $n_e$ simultaneously.

## 4. Results and discussion

### 4.1 The effect of calibration point selection

In this section, the influence of discharge frequency on calibration parameters is studied firstly. In experiment, gas pressure is fixed at 1.5 Pa, Langmuir probe and OES are measured simultaneously, in SF-CCPs. Figure 2(a)-(f) shows the $T_e$ and $n_e$ measured by the Langmuir probe, and obtained by OES with CRM using calibration point of different frequency at the same power. The squares represent the results measure by the Langmuir probe, circles represent the results by OES with CRM using 2 MHz 50 W point calibrated. It means that $T_e$, $n_e$ measured by probe and $I_{750}$, $I_{696}$ measured by OES at 2 MHz 50W discharge, then using the above method to obtain $C_{750}$ and $C_{696}$ values. Furthermore, this curve is using the same $C_{750}$, $C_{696}$ while $I_{750}$, $I_{696}$ for each power to calculate $T_e$, $n_e$ with CRM. Also, triangles represent the results by OES with CRM using 13.56 MHz 50W point calibrated, inverted triangles represent the results by OES with CRM calibrated by 27.12 MHz 50 W point, respectively. Figure 2(a) and (b) are the results in 2 MHz SF-CCPs with power from 50 W to 120 W, since the discharge is unable to sustain when power is below 50 W at this frequency. The $n_e$ increases linearly with the increase of input power, while the $T_e$ decreases with the increase of input power. And the $T_e$ decrease is sharp at first and then it becomes relatively flat.

As shown in figure 2(a), when the three different points with the same power but different frequencies are used to calibrate, the maximum error of $n_e$ between the calculation results from CRM with OES and the probe measurement results is 23.9%, and the average error is 10.7%. The maximum error of $T_e$ between the calculation results from CRM with OES and the probe measurement results is 7.2%, and the average error is 2.6%, as shown in figure 2(b).

Figure 2(c) and (d) are the results in 13.56 MHz SF-CCPs with power from 20 W to 120 W. The maximum error of $n_e$ between the calculation results from CRM with OES and the probe measurement results is 11.9%, and the average error is 6.6%. The maximum error of $T_e$ is 6.1%, and the average error is 2.5%. Figure 2(e) and (f) show the results in 27.12 MHz single-frequency CCPs with power from 20 W to 120 W. The maximum error of $n_e$ between the calculation results by CRM with OES and the probe measurement results is 14.1%, and the average error is 7.8%. The maximum error of $T_e$ is 5.0%, and the average error is 2.4%.

It can be seen when the three different points with the same power but different frequencies are used as punctuation marks, the maximum error of $n_e$ is not more than 25%, and most average error of $n_e$ is less than 10%. The maximum error of $T_e$ is not more than 10%, and the average error of $T_e$ is less than 3%. Therefore, calibration of different frequency points has a limited effect on the results.

Figure 3(a) and (b) show the $n_e$ and $T_e$ measured by the Langmuir probe, and obtained by OES with CRM using calibration point under different input power but same frequency of 13.56 MHz. The squares represent the results measured by the Langmuir probe, the circles represent the results from OES with CRM calibrated by 13.56 MHz 50 W point, the triangles represent the results from OES with CRM using 13.56 MHz 70 W point calibration, and the inverted triangles represent the results from OES with CRM calibrated by 13.56 MHz 90 W point, respectively.

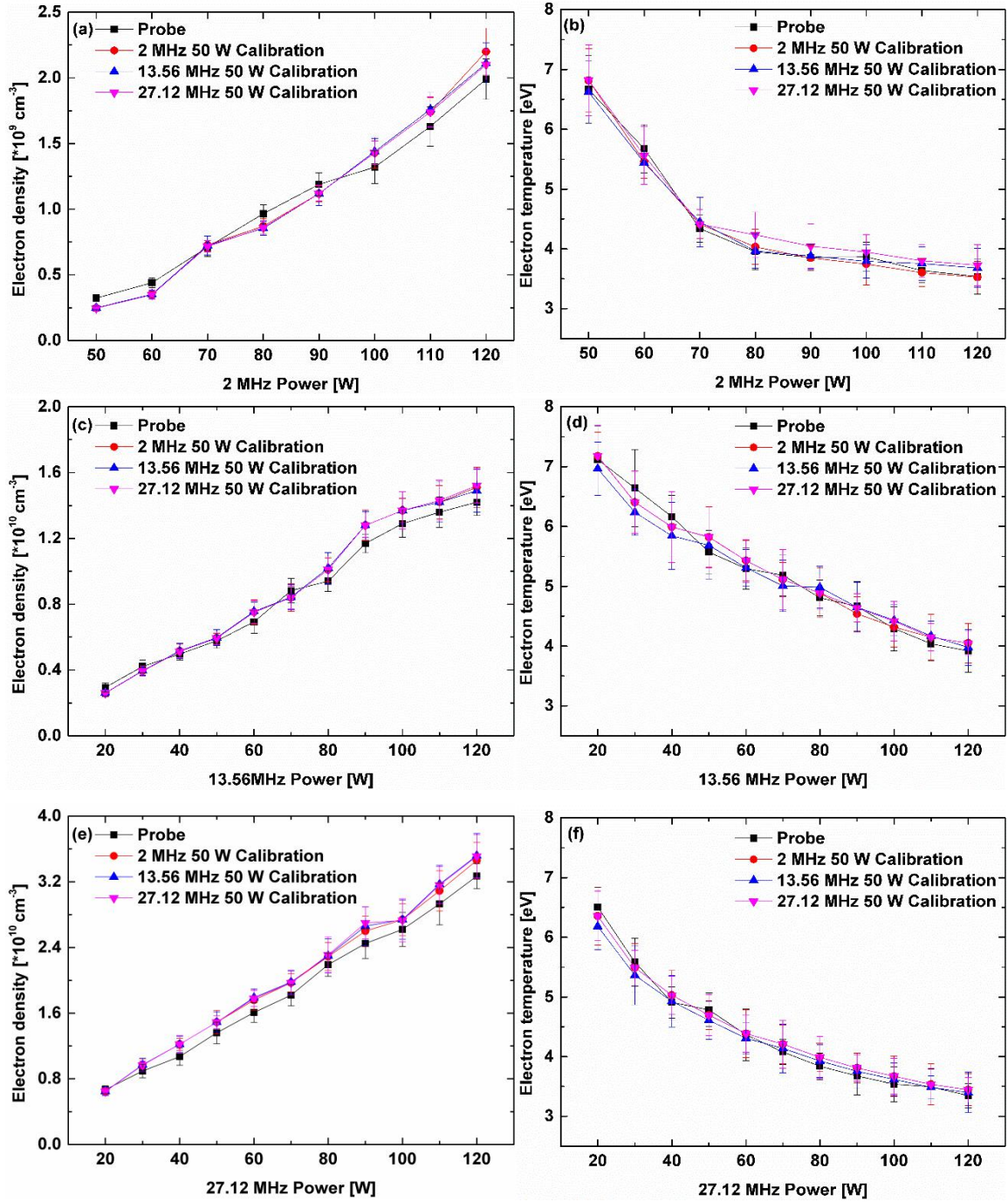

**Figure 2.** The results of (a) electron density and (b) electron temperature from CRM with OES method by using different frequency calibration points as a function of the RF power in SF-CCPs. Note that the squares represent the $n_e$ and $T_e$ measured by Langmuir probe.

As shown in the figure 3(a) and (b), the maximum error of $n_e$ between the calculation results and the probe measurement results is within 15.3%, and the average error of $n_e$ is 6.1%. The maximum error of $T_e$ does not exceed 10.2%, and the average error of $T_e$ is 3.8%. Therefore, calibration of different frequency points has a limited influence on the results.

Hence, at the same gas pressure, any calibration point can be selected to calculate the $n_e$ and $T_e$ in various CCPs, including multi frequency discharges. It means that this method can be used to obtain the $n_e$ and $T_e$ in discharges with the complex radio frequency interference, so as to realize the diagnosis of multi frequency CCPs.

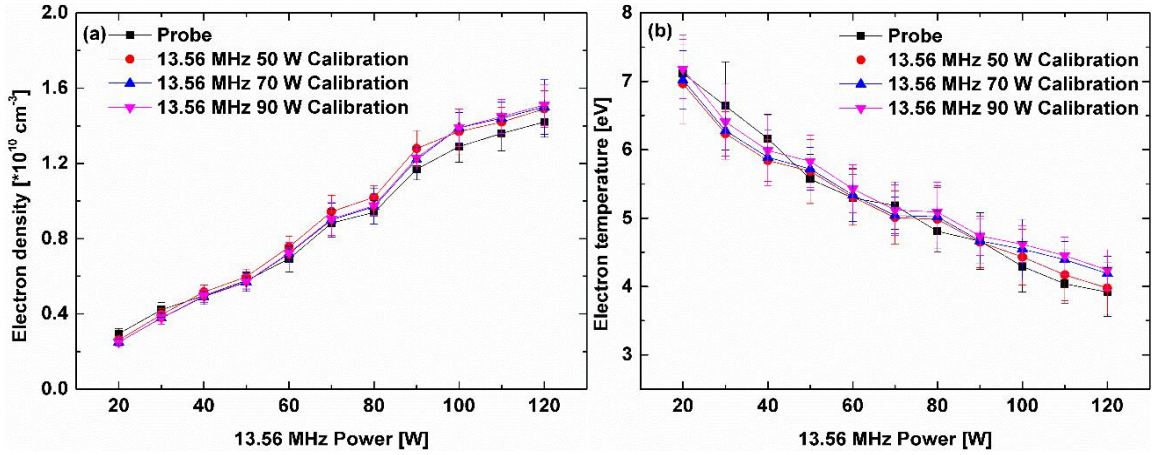

**Figure 3.** The results of (a) electron density and (b) electron temperature from CRM with OES method by using different power calibration points as a function of the RF power in SF-CCPs. Note that the squares represent the $n_e$ and $T_e$ measured by Langmuir probe.

*4.2 The effect of high, medium and low frequency power in TF-CCPs*

In this section, the $n_e$ and $T_e$ are obtained with high, intermediate and low frequency power in TF-CCPs, and the influence of each RF power on the electron temperature and density are discussed. Figure 4(a) and (b) show the changes of $n_e$ and $T_e$ with one RF power of HF (27.12 MHz), IF (13.56 MHz) or LF (2 MHz). The squares represent the HF power changes from 10 W to 120 W, while the remaining two RF power sources are fixed at the power of 5 0W. The circles represent the IF power changes, the other two power are fixed at 50 W and the triangles represent the LF power changes, the other two power are fixed at 50 W, respectively.

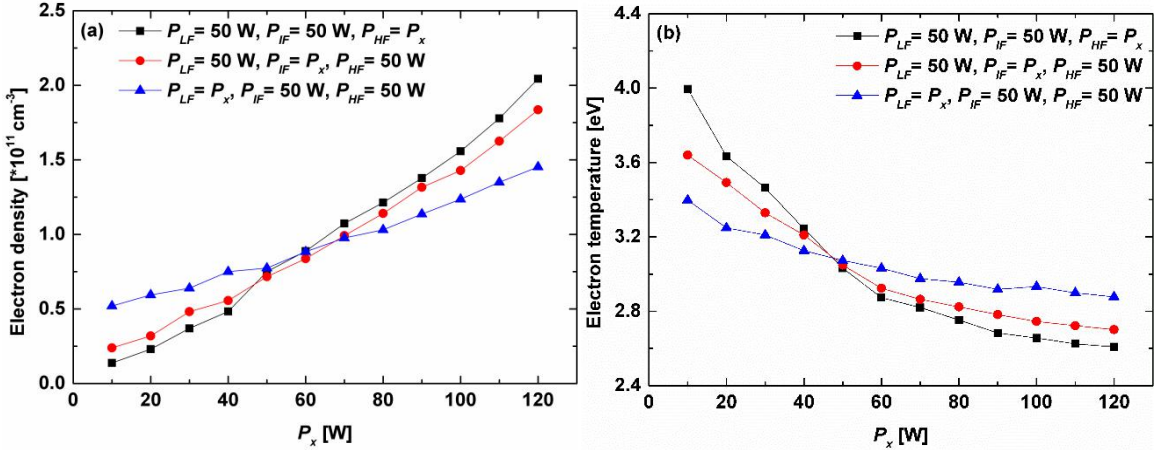

**Figure 4.** The (a) electron density and (b) electron temperature as a function of the HF, IF or LF power in TF-CCPs, respectively. The squares represent the $n_e$ and $T_e$, when LF and IF power are fixed at 50 W. The circles represent the $n_e$ and $T_e$, when LF and HF power are fixed at 50 W. The triangles represent the $n_e$ and $T_e$, when IF and HF power are fixed at 50W.

As shown in the figure 4(a), in the main, $n_e$ increases with HF, IF or LF power in TF-CCPs respectively, while the slopes are different. The change of $n_e$ with HF power is the largest, followed by IF power and LF power. It means that HF power has the greatest effect on $n_e$, followed by IF power, and the least is LF power. When changing the HF power, the IF and LF power are fixed at 50 W. At first, because the proportion of HF power is low in total power, mainly affected by the IF and LF power, the $n_e$ is low. Then as the HF power goes higher than 50 W, the proportion of HF power becomes higher. As a result, mainly affected by the HF power, the $n_e$ increases faster. On the contrary, when changing the LF power, the $n_e$ is higher at first, then the growth rate of $n_e$ becomes slower as the proportion of LF power increases. The $n_e$ curve presented when changing the IF power is in-between that of with the HF power and LF power. It is note that when the power of HF, IF and LF are all 50 W, the $n_e$ curves intersect at a point.

The trend of $T_e$ is opposite to $n_e$, they decrease with HF, IF or LF power in TF-CCPs as shown in the figure 4(b). When the power of HF, IF and LF are all 50 W, the $T_e$ curves also intersect at a point. The change of $T_e$ with HF power also is the biggest, followed by IF power and LF power. It means that HF power has the greatest effect on $T_e$, followed by IF power, and the least is LF power. This is because when the $n_e$ increases, the collision intensifies and the $T_e$ drops. HF power mainly controls $n_e$, therefore, $n_e$ changes most with the increase of HF power, and it leads $T_e$ to vary most with HF power. In fact, the HF, IF and LF are not completely decoupled, while HF power has the greatest influence on $n_e$ and $T_e$, that of LF power is the least, and IF power is in-between.

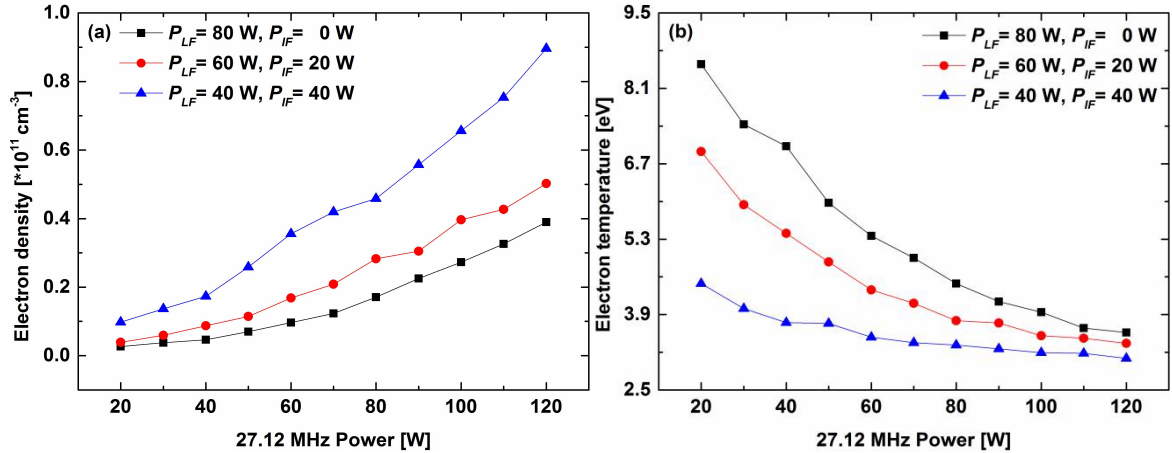

**Figure 5.** The (a) electron density and (b) electron temperature as a function of the HF power when the total of IF and LF power is kept constant, 80 W. The squares represent the $n_e$ and $T_e$ in DF-CCPs, when LF power is fixed at 80 W (IF power is turned off). The circles represent the $n_e$ and $T_e$ in TF-CCPs, when IF power is fixed at 20 W and LF power fixed at 60 W. The triangles represent the $n_e$ and $T_e$ in TF-CCPs, when IF power is fixed at 40W and LF power fixed at 40 W, respectively.

The effect of IF power is demonstrated by comparing different TF discharges and DF discharges. Because IF and LF power source are connected to the one electrode, we keep the total power of IF and LF unchanged. In figure 5, HF power is changed from 20 W to 120 W, the squares represent the $n_e$ and $T_e$ in DF-CCPs, when LF power is fixed at 80 W. (IF power is turned off) The circles represent the $n_e$ and $T_e$ in TF-CCPs, when IF power is fixed at 20 W and LF power at 60 W. And the triangles represent the $n_e$ and $T_e$ in TF-CCPs, when IF power is fixed at 40 W and LF power at 40 W respectively. It means that the total of IF and LF power stays at 80 W.

When the total input power of IF and LF remains constant, increasing the proportion of IF power can increase the electron density, and it becomes more obvious with the increase of the HF power, as shown in figure 5(a). On the contrary, increasing the IF power can decrease the electron temperature, which is more obvious when the HF power is low in figure 5(b). The IF power was used to replace part of the LF power, it can make $n_e$ increase and $T_e$ reduce, and make $T_e$ stable at a lower HF power at the same time. It can be seen from the experimental results that although HF can indeed control the $n_e$ in DF-CCPs, the influence of HF on $n_e$ is strengthened and the influence on $T_e$ is weakened after adding IF power source. The increase of IF can make HF control $n_e$ and $T_e$ more independently.

## 5. Conclusion

In conclusion, an Ar 18-level CRM was established for low-pressure discharges and calibrated by probe. The $n_e$ and $T_e$ in the plasmas can be obtained by combining the OES with the model. Under constant gas pressure, any calibration point can be selected to calculate $n_e$ and $T_e$ in various CCPs. Hence, a method for diagnosing $n_e$ and $T_e$ in discharges with the complex radio frequency interference was developed. Then the $n_e$ and $T_e$ are obtained with high, intermediate and low frequency power in TF-CCPs by this method. It is found that the $n_e$ increases with HF, IF or LF power in TF-CCPs respectively, while the slopes are different, the trend of $T_e$ is opposite to $n_e$. HF power has the greatest influence on $n_e$ and $T_e$, that of LF power is the least, and IF power is in-between. Adding IF power source, the influence of HF on $n_e$ is strengthened and the influence on $T_e$ is weakened. The increase of IF can make HF control $n_e$ and $T_e$ more independently.

## Acknowledgements

This work has been financially supported by the Fundamental Research Funds for the Central Universities (Grant No. 2232020G-10).

## ORCID iD

Xiao-jiang Huang iD https://orcid.org/0000-0003-0320-4773